\documentclass[aps,twocolumn,superscriptaddress]{revtex4}

\usepackage{graphicx}
\usepackage{amsmath}

\begin{document}

\title{Soft x-ray magnetic circular dichroism study of weakly 
ferromagnetic Zn$_{1-x}$V$_x$O thin film}

\author{Y.~Ishida}
\altaffiliation[Present address: ]{RIKEN SPring-8 Center, 
Sayo-cho, Sayo-gun, Hyogo 679-5148, Japan }
\affiliation{Department of Physics and Department of Complexity Science, 
University of Tokyo, Kashiwa, Chiba 277-8561, Japan}

\author{J.I.~Hwang}
\author{M.~Kobayashi}
\affiliation{Department of Physics and Department of Complexity Science, 
University of Tokyo, Kashiwa, Chiba 277-8561, Japan}

\author{Y.~Takeda}
\affiliation{Synchrotron Radiation Research Unit, 
Japan Atomic Energy Agency, 
Sayo, Hyogo 679-5148, Japan}

\author{K.~Mamiya}
\affiliation{Synchrotron Radiation Research Unit, 
Japan Atomic Energy Agency, 
Sayo, Hyogo 679-5148, Japan}

\author{J.~Okamoto}
\altaffiliation[Present address: ]{National Synchrotron 
Radiation Research Center, Hsinchu Science Park, Hsinchu 30076, Taiwan}
\affiliation{Synchrotron Radiation Research Unit, 
Japan Atomic Energy Agency, 
Sayo, Hyogo 679-5148, Japan}

\author{S.-I.~Fujimori}
\author{T.~Okane}
\author{K.~Terai}
\author{Y.~Saitoh}
\affiliation{Synchrotron Radiation Research Unit, 
Japan Atomic Energy Agency, 
Sayo, Hyogo 679-5148, Japan}

\author{Y.~Muramatsu}
\altaffiliation[Present address: ]{Analytical Science Research Group, 
Department of Materials Science and Chemistry, 
Graduate School of Engineering, University of Hyogo, 
Himeji, Hyogo 671-2201, Japan}
\affiliation{Synchrotron Radiation Research Unit, 
Japan Atomic Energy Agency, 
Sayo, Hyogo 679-5148, Japan}

\author{A.~Fujimori}
\affiliation{Department of Physics and Department of Complexity Science, 
University of Tokyo, Kashiwa, Chiba 277-8561, Japan}
\affiliation{Synchrotron Radiation Research Unit, 
Japan Atomic Energy Agency, 
Sayo, Hyogo 679-5148, Japan}

\author{A.~Tanaka}
\affiliation{Graduate School of Advanced Sciences of Matter, 
Hiroshima University, Higashi-Hiroshima, Hiroshima 739-8530, Japan
}

\author{H.~Saeki}
\author{T.~Kawai}
\author{H.~Tabata}
\affiliation{The Institute of Scientific and Industrial Research, 
Osaka University, Ibaraki, Osaka 567-0047, Japan}

\date{\today}

\begin{abstract}
We performed a soft x-ray magnetic circular dichroism (XMCD) 
study of a Zn$_{1-x}$V$_x$O thin film which showed 
small ferromagnetic moment. Field and temperature dependences of V 2$p$ 
XMCD signals 
indicated the coexistence of Curie-Weiss paramagnetic, antiferromagnetic, 
and possibly ferromagnetic V ions, 
quantitatively consistent with the magnetization measurements. 
We attribute the paramagnetic signal to V ions substituting 
Zn sites which are somewhat elongated along the $c$-axis. 
\end{abstract}


\maketitle

Recently, ZnO-based diluted magnetic semiconductors (DMSs) have 
attracted much interest due to their potentially 
high Curie temperatures ($T_{\rm C}$'s) \cite{Dietl}
aiming at practical applications in spintronics devices \cite{DasSarma}. 
After the report of high $T_{\rm C}$ ($\sim$280\,K) in 
Zn$_{1-x}$Co$_x$O \cite{UedaZnCoO}, 
there have been many reports on 
ZnO-based DMSs showing high $T_{\rm C}$'s 
(for recent review, see \cite{Liu}). 
Zn$_{1-x}$V$_{x}$O (ZVO) thin films prepared by the pulsed laser 
deposition method under a reduced atmospheric condition 
showed $T_{\rm C}$'s above 
400\,K \cite{Saeki1}. The ferromagnetism in ZVO 
was reproduced in \cite{Venkatesan, Neal}, but 
no ferromagnetic behavior was observed in 
a recent report \cite{NoFerro}. 
Thus, the high $T_{\rm C}$ in ZVO has remained 
controversial concerning its origin \cite{Coey}. 
In order to exclude possible metal precipitations as an extrinsic origin of 
ferromagnetism, 
magnetization measurements or 
anomalous Hall effect measurements may not be sufficient \cite{ZnCoOHall}. 
In this Letter, we report on a soft x-ray magnetic circular dichroism (XMCD) 
study of ZVO. 
XMCD in core-level soft x-ray absorption spectroscopy (XAS) is an 
element specific probe and is sensitive to the magnetic states of 
each element. 
Their line shapes are fingerprints of 
the electronic structures such 
as the valence and the crystal field of the magnetic 
ion, making XMCD a powerful tool to 
investigate the electronic and magnetic properties of DMSs 
\cite{Ohldag, Ueda, Keavney, Edmonds, Rader, Masaki, KeavneyII}. 

A thin film of ZVO ($x$=0.05) was epitaxially grown on a 
ZnO(0001) buffer layer on an Al$_2$O$_3$(1120) substrate 
as described elsewhere \cite{Saeki1}. 
The film size was approximately 5 mm$\times$7 mm with the thickness of 100\,nm. 
XAS and XMCD measurements were performed at 
BL23-SU of SPring-8 \cite{Okamoto_23SU}. 
The monochromator 
resolution was $E$$/$$\varDelta$$E$$>$10000.
The photon helicity ($>$90\,\% circular polarization within the $ab$-plane) 
was switched 
at each photon energy. Magnetic field $H$ was applied 
parallel and antiparallel to the $c$-axis. 
Sample surface was cleaned {\it in situ} by Ar-ion sputtering at 1\,keV 
and subsequent annealing up to 200$^{\circ}$C. 
We also performed magnetization measurements after the 
XAS and XMCD measurements using a SQUID magnetometer (MPMS, Quantum 
Design Co., Ltd).

\begin{figure}[htb]
\begin{center}
\includegraphics[width=8.7cm]{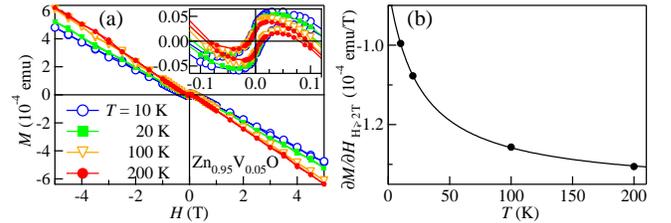}
\caption{\label{Mag}(Color online) Results of magnetization 
measurements on Zn$_{0.95}$V$_{0.05}$O. 
(a) Magnetization curves at 
various temperatures. Inset shows a magnified plot near $H=0$. 
(b) High-field susceptibility (see, text) {\it vs} temperature fitted to 
the Curie-Weiss law (solid curve), $\partial M/\partial H_{H > 2\,{\rm T}} = 
NC/(T - \varTheta) + \partial M/\partial H_{\rm 0}$, where 
$\partial M/\partial H_{\rm 0} = -(1.361\pm 0.010)\times 10^{-4}$\,emu/T, 
$NC = (1.30\pm 0.16)\times 10^{-3}$\, emu\,K/T, 
and $\varTheta = -25.4\pm 3.6$\,K. }
\end{center}
\end{figure}

Figure \ref{Mag}(a) shows magnetization curves of the ZVO film 
taken at various temperatures. We observed hysteresis loops 
on a strong diamagnetic background of the substrate and confirmed 
that the sample was ferromagnetic with $T_{\rm C} > 200$\,K [see 
inset of Fig.\ \ref{Mag}(a)]. 
The ferromagnetic moment was $\sim$6$\times 10^{-3}$\,$\mu_{\rm B}$/V ion. 
In Fig.\ \ref{Mag}(b), we have plotted a high-field ($H >$ 2\,T) 
magnetic susceptibility, 
$\partial M/\partial H_{H > 2\,{\rm T}}$, 
as a function of $T$, where 
$M$ is the magnetization of the sample. 
$\partial M/\partial H_{H > 2\,{\rm T}}$ was 
fitted to the Curie-Weiss (CW) law 
with an offset, 
	$\partial M/\partial H_{H > 2\,{\rm T}} = 
	NC/(T - \varTheta) + \partial M/\partial H_{\rm 0}$, 
where 
	$C = (g\mu_B)^2S(S+1)/3k_B$ is the Curie constant, 
	$\varTheta$ is the Weiss temperature, 
	$\partial M/\partial H_{\rm 0}$ is a constant, 
	$N$ is the number of magnetic ions in the sample, 
    and $g$ is the $g$-factor. 
$\partial M/\partial H_{\rm 0}$ contains the diamagnetic and 
temperature independent paramagnetic contribution. 
The excellent fit indicates that the temperature dependence of 
$\partial M/\partial H_{H > 2\,{\rm T}}$ is caused by 
the local magnetic moments with antiferromagnetic correlations. We attribute 
this CW behavior to those of the V$^{2+}$ ions (described below). 
Assuming $g=2$ and $S=$ 3/2, 
one obtains $N = 0.41$$\times$$10^{15}$, which is estimated to be 
$\sim$10\,\% of the total V atoms in the sample.

\begin{figure}[htb]
\begin{center}
\includegraphics[width=7cm]{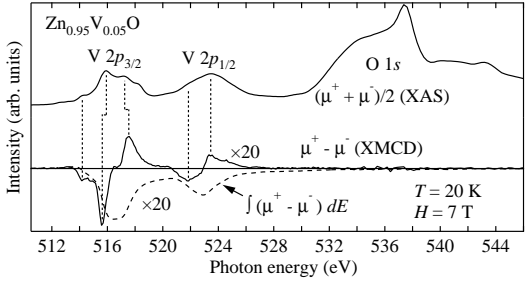}
\caption{\label{XMCD_1}XAS and XMCD at the V 2$p$ and O 1$s$ 
absorption edges of ZVO ($x=0.05$) recorded at 
$T=20$\,K and $H=7$\,T.}
\end{center}
\end{figure}

Figure \ref{XMCD_1} shows V 2$p$ and O 1$s$ XAS and 
XMCD taken at $T=$ 20\,K under $H=7$\,T. 
The structures around $h\nu =$ 516\,eV, 524\,eV, and $>$530\,eV 
are the V 2$p$$_{3/2}$, 
V 2$p$$_{1/2}$, and O 1$s$ absorption edges, respectively. 
We note that Ar-ion sputtering have reduced the peak at 518.3\,eV 
(most likely due to contamination in the surface region), 
although it was not completely removed. 
The O 1$s$ XAS showed a sharp peak at $h\nu=$ 537.5\,eV 
and a plateau at $h\nu\sim$ 540 - 543\,eV, similar 
to that of highly oriented ZnO microrod arrays 
with polarization vector within the $a$$b$-plane \cite{ZnOXAS}. 
This indicates that the $c$-axis of ZVO was 
oriented to surface normal. 
The V 2$p$ XAS and XMCD show multiplet structures, indicating that the 
doped V atoms were in an oxidized state and not 
in a metallic state such as metallic clusters. 
The strongest negative and positive XMCD signals were observed at 
$h\nu=$ 515.6\,eV and 517.5\,eV, respectively, which were 
different from the peak positions of $h\nu=$ 515.9\,eV and 517.3\,eV 
in the V 2$p$ XAS spectrum. 
This may be explained in a magnetically inhomogeneous picture of the 
V ions that there exists an XMCD-active minority component whose electronic 
environment differs from the XMCD-inactive majority component. 
The energy integral of 
the V 2$p$ XMCD signal 
was close to zero (or even slightly negative: see, Fig.\ \ref{XMCD_1}), 
indicating that the orbital 
magnetic moment of the V 3$d$ electrons 
is quenched from the ionic value \cite{Thole}.

\begin{figure}[htb]
\begin{center}
\includegraphics[width=7.7cm]{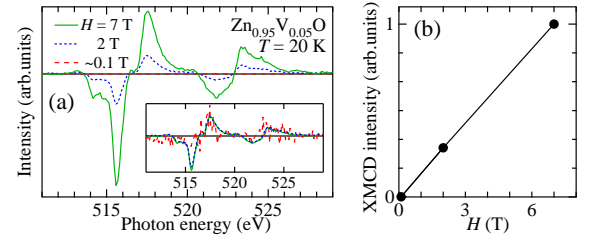}
\caption{\label{Field}(Color online) Magnetic field dependence of 
XMCD at the V 2$p$ absorption edge. (a) Raw spectra taken at 
$T=$ 20\,K. Inset shows the normalized spectra. (b) XMCD signal intensity 
{\it vs} magnetic field. }
\end{center}
\end{figure}

Figure \ref{Field}(a) shows the $H$ dependence of 
V 2$p$ XMCD at 20\,K. 
Since XMCD taken at $\sim$$0.1$\,T was small on 
the scale of Fig.\ \ref{Field} \cite{footnote1}, we show normalized 
XMCD in the inset in Fig.\ \ref{Field}(a). 
Figure \ref{Field}(b) shows the XMCD intensity 
as a function of $H$. The linear increase of the 
XMCD signal with $H$ indicates that the paramagnetic signal dominates the 
XMCD signal and that the 
ferromagnetic 
component is small, consistent with the magnetization 
measurements. 
The line shapes under 2\,T and 
7\,T were nearly identical [inset in Fig.\ \ref{Field}(a)]. 
The V 2$p$ XMCD signals at 170\,K under 2\,T was 
$\sim$30\,\% \cite{footnote2} 
of that at 20\,K under 2\,T. 
This indicates that the paramagnetism observed in the V 2$p$ 
XMCD signal is mainly reflecting the CW paramagnetism of the 
V ions. In fact, 
using $\varTheta=25$\,K derived from the magnetization measurements 
(Fig.\ \ref{Mag}), the CW susceptibility 
at 170\,K becomes 23\,\% of that at 20\,K, 
which explains most of the XMCD intensity decrease from 
20\,K to 170\,K. The majority of the V ions were 
presumably strongly coupled antiferromagnetically 
(N{\'e}el temperature $\gtrsim$1000\,K \cite{Masaki}), 
and hence its contribution 
to the susceptibility was negligible.

\begin{figure}[htb]
\begin{center}
\includegraphics[width=6cm]{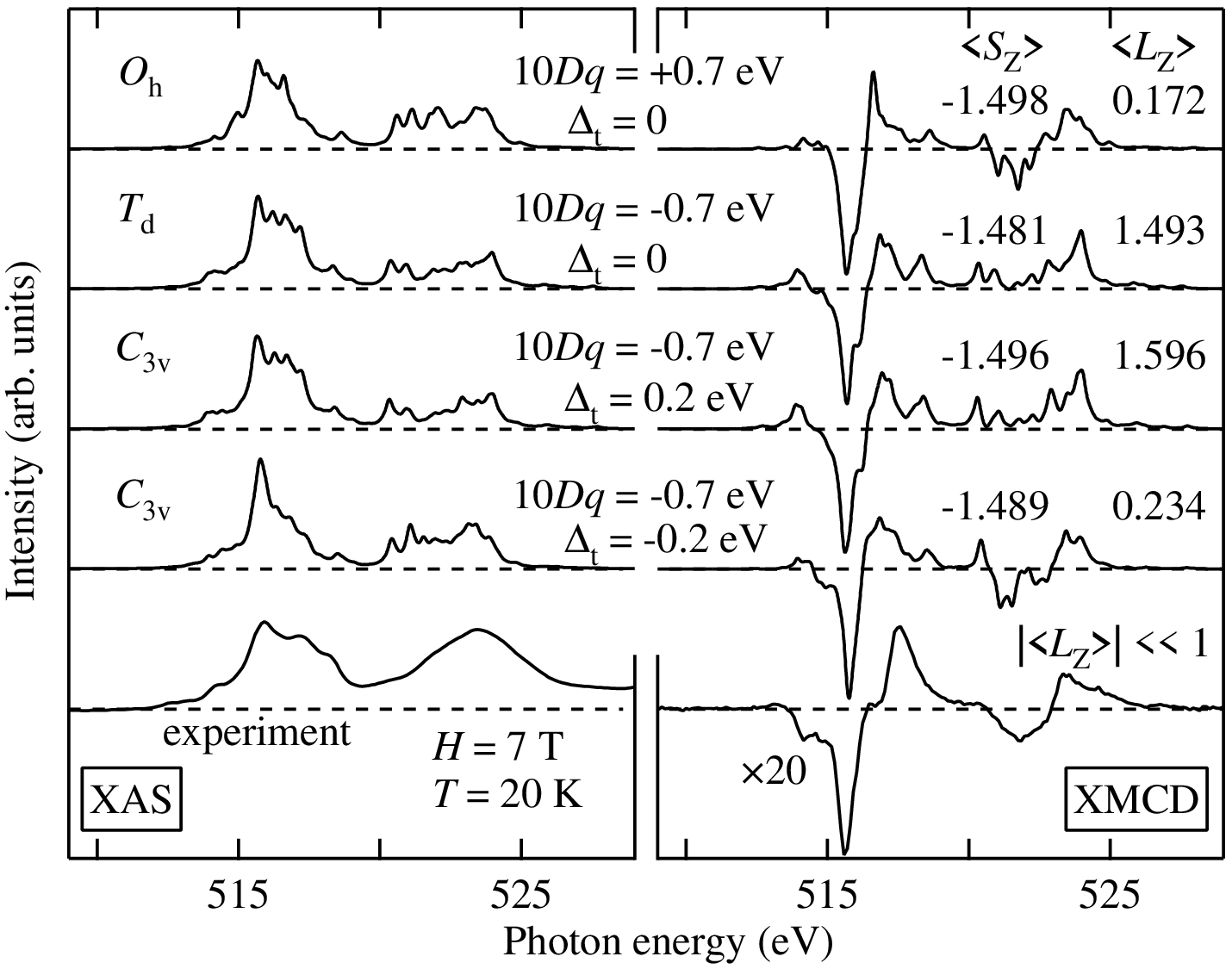}
\includegraphics[width=2.5cm]{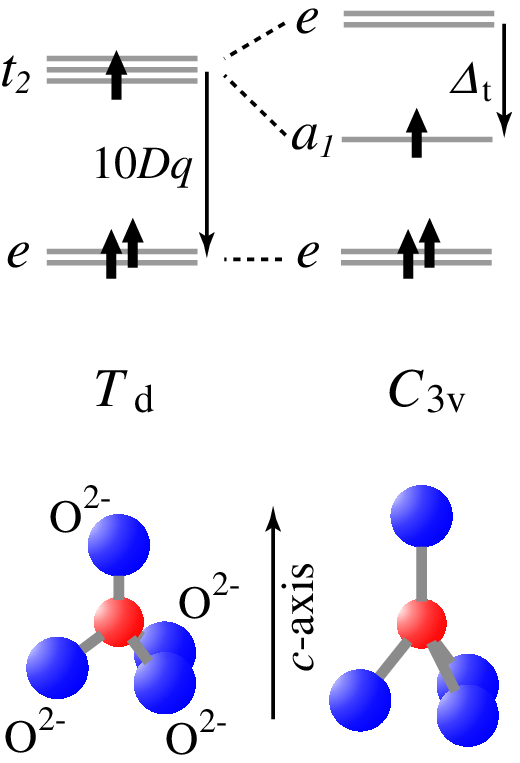}
\caption{\label{Ion}(Color online) Comparison of V 2$p$ XAS and XMCD with 
atomic multiplet calculations of V$^{2+}$ ions. 
$\langle S_z\rangle$ and $\langle L_z\rangle$ (in unit of $\mu_B$) 
are also shown. Crystal field parameters are schematically shown in the 
right panel.}
\end{center}
\end{figure}

In Fig.\ \ref{Ion}, we show 
atomic multiplet calculations for V$^{2+}$ \cite{footnote3} under 
octahedral ($O_{\rm h}$), tetrahedral ($T_{\rm d}$), and 
trigonal ($C_{\rm 3v}$) symmetries by introducing the 
crystal field parameters 10$Dq$ and $\varDelta_{\rm t}$ (see right panel 
in Fig.\ \ref{Ion}). 
$C_{\rm 3v}$ symmetry arises from a 
slight elongation or contraction of the tetrahedron 
along the $c$-axis of ZnO having the wurtzite structure \cite{ZnO_crystal}. 
The orbital and spin magnetic moments, 
$\langle L_z\rangle$ and $\langle S_z\rangle$, are 
also indicated in the figure. 
Large $\langle L_z\rangle$ is present in the case of $T_{\rm d}$ or 
$C_{\rm 3v}$ with $\varDelta_{\rm t} < 0$ 
while those in the case of $O_{\rm h}$ or $C_{\rm 3v}$ with 
$\varDelta_{\rm t} > 0$ are small, 
because the orbital degeneracy remains for $T_{\rm d}$ and 
$C_{\rm 3v}$ with $\varDelta_{\rm t} < 0$ 
while it is lifted for $O_{\rm h}$ and $C_{\rm 3v}$ with 
$\varDelta_{\rm t} > 0$. Since 
the V 2$p$ XMCD indicated a quenched $\langle L_z\rangle$, 
we could exclude 
$T_{\rm d}$ and $C_{\rm 3v}$ with $\varDelta_{\rm t} < 0$. 
A shoulder structure around $h\nu\sim$ 514\,eV in the XMCD 
was reproduced in $C_{\rm 3v}$ with $\varDelta_{\rm t} > 0$. 
Thus, we conclude that the V 2$p$ XMCD 
came from substitutional V$^{2+}$ ions under 
slight elongation of the 
tetrahedra along the $c$-axis of ZnO \cite{ZnO_crystal}. 
On the other hand, the experimental V 2$p$ XAS line shape 
resembled the spectra for $T_{\rm d}$ or $C_{\rm 3v}$ 
with $\varDelta_{\rm t} < 0$. 
Therefore, the majority of the V ions were expected to be 
at the substitutional sites as divalent ions 
without significant elongation of the tetrahedra (reduced $\varDelta_{\rm t}$).

The CW paramagnetic magnetization of the sample 
at 20\,K under 2\,T (Fig.\ \ref{Mag}) could be 
explained by $\sim$2\,\% 
of the full magnetization of the 
V$^{2+}$ ions. 
This value was in good agreement with the observed V 2$p$ XMCD 
intensity which was $\sim$5\,\% of that of the atomic multiplet calculation 
(note that the experimental XMCD is magnified by a factor of 20 
in Fig.\ \ref{Ion}). 
Thus, we could associate the CW paramagnetic behavior in 
$\partial M/\partial H_{H>2\,{\rm T}}$ to the CW paramagnetic behavior 
in V 2$p$ XMCD and hence to the substitutional V$^{2+}$ under 
slight elongation along the $c$-axis. 

In the previous XMCD study of Zn$_{1-x}$Co$_x$O \cite{Masaki}, 
the ferromagnetic and paramagnetic Co ions were found to exist 
in similar electronic environment \cite{Masaki}. Therefore, although 
not observed in this study, 
the ferromagnetic V ions may be in 
a similar environment to that of the paramagnetic V ions identified in this study. 
It may be possible that 
subtle environmental differences, e.g., in the nearest 
neighbor cations and/or neighboring defects 
determine whether the magnetic ion becomes ferromagnetic or 
paramagnetic in transition-metal-doped ZnO. 

In summary, the field and temperature dependence of the 
V 2$p$ XMCD of ZVO ($x=$ 0.05) showing small ferromagnetic 
moment revealed that $\sim$10\,\% of the V ions were 
CW paramagnetic, $\sim$90\,\% were presumably 
strongly coupled antiferromagnetically, and the 
ferromagnetic component was below 
the detection limit of XMCD. Elongation along the ZnO $c$-axis was 
important in order to explain the XMCD line shape and 
the quenched orbital magnetic moment of 
the V$^{2+}$ CW paramagnetic component. 
Our study suggests that local lattice distortion and subsequent 
orbital anisotropy are important 
in explaining the magnetism of ZnO-based DMSs. 

The magnetization measurements were performed 
at the Materials Design and Characterization 
Laboratory, Institute for Solid State Physics, University of Tokyo. 
This work was supported by a Grant-in-Aid for Scientific 
Research in PriorityArea 
``Invention of Anomalous Quantum Materials" (16076208) 
from MEXT, Japan.

\bibliographystyle{apasoft}

\begin{thebibliography}{99}

\bibitem{Dietl}
T. Dietl {\it et~al.}, Science {\bf 287},  1019  (2000).

\bibitem{DasSarma}
I. \v{Z}uti{\' c}, J. Fabian, and S.~D. Sarma, Rev. Mod. Phys. {\bf 76},  323  (2004).

\bibitem{UedaZnCoO}
K. Ueda, H. Tabata, and T. Kawai, Appl. Phys. Lett. {\bf 79},  988  (2001).

\bibitem{Liu}
C. Liu, F. Yun, and H. Morkoc, J. Mater. Sci: Mat. Electronics {\bf 16},  555
  (2005).

\bibitem{Saeki1}
H. Saeki, H. Tabata, and T. Kawai, Solid State Commun. {\bf 120},  439  (2001).

\bibitem{Venkatesan}
M. Venkatesan, C.~B. Fitzgerald, J.~G. Lunney, and J.~M.~D. Coey, Phys. Rev.
  Lett. {\bf 93},  177206  (2004).

\bibitem{Neal}
J.~R. Neal {\it et~al.}, Phys. Rev. Lett. {\bf 96},   197208   (2006).

\bibitem{NoFerro}
S. Ramachandran, A. Tiwari, J. Narayan, and J.~T. Prater, Appl. Phys. Lett.
  {\bf 87},  172502  (2005).

\bibitem{Coey}
J.~M.~D. Coey, M. Venkatesan, and C.~B. Fitzgerald, Nature mater.\ {\bf 4},
  173  (2005).

\bibitem{ZnCoOHall}
S.~R. Shinde {\it et~al.}, Phys. Rev. Lett. {\bf 92},  166601  (2004).

\bibitem{Ohldag}
H. Ohldag {\it et~al.}, Appl. Phys. Lett. {\bf 76},  2928  (2000).

\bibitem{Ueda}
S. Ueda {\it et~al.}, Physica E {\bf 10},  210  (2001).

\bibitem{Keavney}
D.~J. Keavney {\it et~al.}, Phys. Rev. Lett. {\bf 91},  187203  (2003).

\bibitem{Edmonds}
K.~W. Edmonds {\it et~al.}, Phys. Rev. B {\bf 71},  064418  (2005).

\bibitem{Rader}
O. Rader {\it et~al.}, J.\ Elec.\ Spec.\ Relat.\ Phenom. {\bf 144-147},  789
  (2005).

\bibitem{Masaki}
M. Kobayashi {\it et~al.}, Phys. Rev. B {\bf 72},  201201  (2005).

\bibitem{KeavneyII}
D.~J. Keavney {\it et~al.}, Phys. Rev. Lett. {\bf 95},  257201  (2005).

\bibitem{Okamoto_23SU}
J. Okamoto {\it et~al.}, AIP Conf.\ Proc.\ {\bf 705},  1110  (2004).

\bibitem{ZnOXAS}
J.~H. Guo {\it et~al.}, J.\ Phys.: Condens. Matter {\bf 14},  6969  (2002).

\bibitem{Thole}
B.~T. Thole, P. Carra, F. Sette, and G. van~der Laan, Phys. Rev. Lett. {\bf
  68},  1943  (1992).

\bibitem{footnote1}
The 
precise control of the small magnetic field 
in the range of 0.1\,T was difficult due to the presence of 
residual magnetization in the superconducting magnet, and hence 
we denote $\sim$0.1\,T, 
as estimated from the current of the superconducting 
magnet.

\bibitem{footnote2}
The XMCD at 170\,K under 2\,T was derived 
only from applying $H$ parallel to the propagation of the incident 
light, and hence the large uncertainty in the estimation of the 
intensity. This does not affect our conclusion of the 
strong suppression of the V 2$p$ XMCD intensity at 
high temperatures. 

\bibitem{footnote3}
The cases for 
V$^{3+}$ was precluded from Fig.\ \ref{Ion} since their line shapes 
poorly agreed with those of the experiment.

\bibitem{ZnO_crystal}
S.~C. Abrahams and J.~L. Bernstein, Acta.\ Cryst.\ {\bf 25},  1233  (1969).

\end{thebibliography}

\end{document}